\documentclass[twocolumn,english,prl]{revtex4}
\usepackage[T1]{fontenc}
\usepackage[latin9]{inputenc}
\usepackage{color}
\usepackage{graphicx}

\makeatletter
\@ifundefined{definecolor}
 {\usepackage{color}}{}
\@ifundefined{definecolor}
 {\@ifundefined{definecolor}
 {\usepackage{color}}{}
}{}
\@ifundefined{definecolor}
 {\@ifundefined{definecolor}
 {\@ifundefined{definecolor}
 {\usepackage{color}}{}
}{}
}{}
\@ifundefined{definecolor}
 {\@ifundefined{definecolor}
 {\@ifundefined{definecolor}
 {\@ifundefined{definecolor}
 {\usepackage{color}}{}
}{}
}{}
}{}

\makeatother

\usepackage{babel}

\makeatother

\usepackage{babel}

\makeatother

\usepackage{babel}

\makeatother

\usepackage{babel}

\begin{document}

\title{Hanbury Brown and Twiss Correlations of Anderson Localized waves}

\author{Y. Lahini$^{1}$, Y. Bromberg$^{1}$, Y. Shechtman$^{2}$, A. Szameit$^{2}$,
\textcolor{black}{D. N. Christodoulides$^{3}$} R. Morandotti$^{4}$
and Y. Silberberg$^{1}$}

\affiliation{$^{1}$Department of Physics of Complex Systems, Weizmann Institute
of Science, Rehovot, Israel.}

\affiliation{$^{2}$Department of Solid State Physics, Technion, Israel}

\affiliation{$^{3}$CREOL/College of Optics, University of Central Florida, Orlando,
Florida, USA}

\affiliation{$^{4}$Institute National de la Recherche Scientifi{}que, Varennes,
Qubec, Canada}
\begin{abstract}
When light waves propagate through disordered photonic lattices, they
can eventually become localized due to multiple scattering effects.
Here we show experimentally that while the evolution and localization
of the photon density distribution is similar in the two cases of
diagonal and off-diagonal disorder, the density-density correlation
carries a distinct signature of the type of disorder. We show that
these differences reflect a symmetry in the spectrum and eigenmodes
that exists in off-diagonally disordered lattices but is absent in
lattices with diagonal disorder.
\end{abstract}
\maketitle
\textcolor{black}{The propagation of quantum-mechanical waves in periodic
and disordered media is a fundamental theme in solid state physics,
underlying the transport properties of condense matter systems. In
a perfectly periodic system, the translational invariance gives rise
to extended eigenmodes known as the Bloch modes. As a result, in periodic
systems an initially narrow wavepacket will expand indefinitely and
ballistically, i.e. its width will grow linearly in time. Disorder
in an o}therwise perfectly periodic lattice breaks the translational
symmetry and can lead to exponential localization of the system's
eigenmodes and to the arrest wavepacket expansion (or diffusion) -
a phenomena known as Anderson localization \cite{Anderson,Lee}.

Traditionally, the localization of waves inside the medium was not
observed directly, but rather inferred indirectly from transmission
or conductance measurements. Recently, a new approach to localization
of light was realized using disordered photonic lattices \cite{Pertch,segev,Lahini,hagai,Boundry_loc},
in which light propagates freely along one axis, and exhibits localization
in the transverse directions ({}``transverse'' localization \cite{TLoc}).
The equations describing the propagation of light in these systems
are identical to the equations describing the evolution of a single
quantum particle in an atomic lattice, under the tight-binding approximation,
thus allowing for the direct observation of Anderson localization
as originally described in \cite{Anderson}. In these experiments,
a localized wavepacket, typically a single site wide, was released
inside the disordered lattice and allowed to expand. In periodic lattices,
such experiments led to the observation of ballistic wavepacket expansion
\cite{RevNature,ReviewRep}. As a result of the disorder the wavepacket
exhibited a modified expansion profile, the features of which depend
on the dimensionality of the system \cite{segev,Lahini}, eventually
settling to an exponentially localized distribution - the hallmark
of Anderson localization. Recently, a similar approach enabled the
direct observation of Anderson localization of matter wave \cite{Billy,Roati,Chabe}
in disordered optical potentials, also described by the same equations
\cite{Modungo}. 

\begin{figure}
\includegraphics[clip]{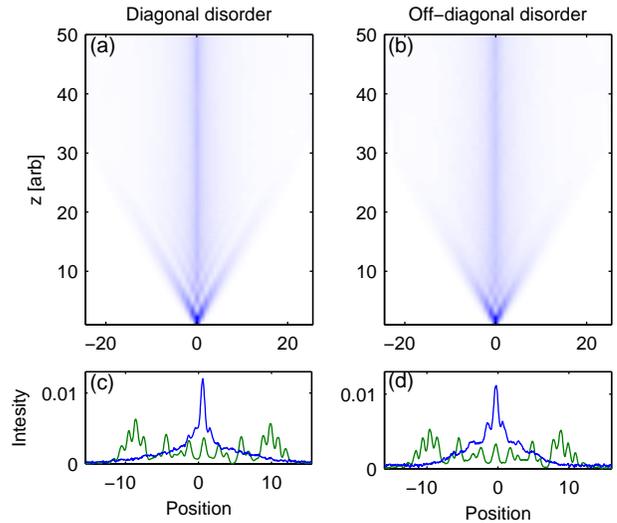}

\caption{\label{fig:1}(color online). Observation of Anderson localization
in lattices with diagonal and off-diagonal disorder. Top panels show
simulation of the average wavepacket expansion in lattices with diagonal
(a) and off-diagonal (b) disorder, when excited at a single site,
showing practically the same dynamics. (c), (d) Experimental measurements
of the output distributions for the two types of disorder (blue lines)
as compared to the same distribution in a periodic lattice (green). }

\end{figure}

While it is well established that in Anderson localization the average
density distribution exhibits exponential localization, not much is
known about higher correlations of the localized fields. Here we show
that spatial intensity correlations, also known as Hanbury Brown and
Twiss correlations \cite{HBT}, do not necessarily display a uniform
decay. Moreover, in contrast to the density distribution, we find
that these correlations carry a signature of the type of disorder:
while light localized in lattices with on-site (diagonal) disorder
show decaying correlations, the correlations in lattices with random
tunneling amplitudes (off-diagonal disorder) exhibit oscillations.
We relate the observed correlation features to a spectral symmetry
that exists in lattices with off-diagonal disorder and is absent in
lattices with diagonal disorder. Specifically, the eigenvalues in
these disordered lattices are anti-symmetrically distributed about
the mean value, and eigenmodes associated with symmetric eigenvalues
share several properties. These results are related to a recent prediction
of quantum correlations in the single particle limit \cite{Qanderson},
and we now show that some analogous features can be also be observed
in the classical regime. While we shall discuss here only the one
dimensional problem, these effects extend to the two and three dimensional
problems as well. 

The description for Anderson localization of light in one dimensional
waveguide lattices is given by set of coupled discrete Schrodinger
equations: \begin{equation}
-i\frac{\partial{U_{n}}}{\partial{z}}=\beta_{n}U_{n}+C_{n,n+1}U_{n+1}+C_{n,n-1}U_{n-1}\end{equation}
 Here $n=1,...,N$ where $N$ is the number of lattice sites (waveguides),
$U_{n}$ is the wave amplitude at site $n$, $\beta_{n}$ is the eigenvalue
(propagation constant) associated with the $n$th site , $C_{n,n\pm1}$
are the tunneling amplitudes between two adjacent sites, and $z$
is the longitudinal space coordinate (for a more detailed description
see e.g. \cite{Lahini}). These equations are identical to the equations
describing the time evolution of a single electron in a lattice under
the tight binding approximation \cite{RevNature,ReviewRep}, where
$z$ represent time, and $U_{n}$ is the wavefunction at site $n$.
Therefore, while the experiments described in this Letter were conducted
in the optical domain, the results hold also for other systems described
by the tight-binding model, such as electron in crystalline structure,
or Bose-Einstein condensate in disordered optical potentials.

In the tight-binding model, the disorder type falls into two broad
categories: diagonal disorder, in which the $\beta_{n}$ parameters
are randomized, but the tunneling amplitudes $C_{n,n\pm1}$ are fixed
across the lattice. Such disorder was considered by Anderson in his
original work \cite{Anderson}, in what is now known as the Anderson
model. With few exceptions \cite{hagai,Boundry_loc}, all the recent
experiments reporting the observation of Anderson localization of
light \cite{segev,Lahini} and matter waves \cite{Billy,Roati} were
conducted using this type of disorder. A second type of theoretically
well studied disorder is known as 'off-diagonal' disorder, in which
the the $\beta_{n}$ parameters are fixed across the lattice, yet
the tunneling amplitudes are randomized. Such lattices are known to
exhibit several unique spectral properties \cite{Dyson,Inui,Soukoulis,Soukoulis-1,Komiyama}.
However, very little has been achieved so far to experimentally observe
a signature of these properties. 

The recent experiments on AL of light \cite{Pertch,segev,Lahini}
and matter waves \textcolor{black}{\cite{Billy,Roati}, have reported
the direct measurement of the main features of localization, namely
the cross-over from ballistic transport to localization as a function
of time and the level of disorder. For example, Fig. 1 shows simulations
of the wavepacket dynamics in disordered one dimensional lattices
\cite{Lahini}. This dynamics starts with a ballistic expansion of
the wavepacket, similarly to the expansion in perfectly periodic lattices.
After some propagation, a localized component emerges near }the\textcolor{black}{
origin, co-existing with the transient, ballistic component. As the
waves propagate, the ballistic component decays and the intensity
distribution becomes exponentially localized in space. Fig. 1 compares
this evolution in lattices with diagonal (a) and off-diagonal (b)
disorders, showing practically identical evolution.}

\begin{figure}
\includegraphics[clip]{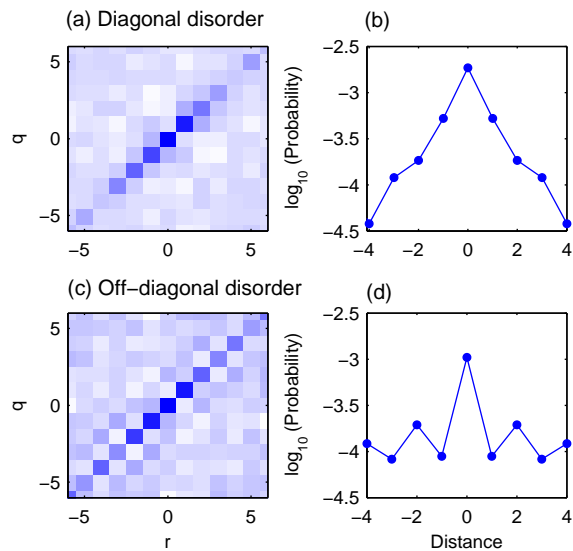}

\caption{\label{fig:2}(color online) Measured density-density correlations
$\Gamma_{r,q}=<I_{r}I_{q}>/<I_{r}><I_{q}>$ for localized wavepackets
in: (a) lattices with diagonal disorder. The strong diagonal feature
reflects finite coherence length of the waves (see text). (b) The
extracted correlation function. (c) Density correlations in lattices
with off-diagonal disorder, showing checker-like correlations. (d)
The extracted correlation function, showing oscillating correlations.}

\end{figure}

We have fabricated two types of disordered lattices of waveguides,
similar to those used to demonstrate Anderson Localization of light
\cite{Lahini}. In one array, the waveguide widths were identical,
but their separations were randomized, modeling off-diagonal disorder.
The other, a diagonally disordered array, had random-width waveguides
yet constant separations. Light was launched into individual waveguides
and the output distribution was recorded. When the output intensity
was averaged over \textasciitilde{}100 realizations (by launching
the light at different input locations) we reproduced the exponentially
decaying localization shown in Figure 1(c) and (d). It is shown in
comparison with the ballistic expansion that was measured in perfectly
periodic arrays. We note that the two types of disorder led to very
similar localized states.

A significant difference was observed, however, when we measured the
density correlations of the output distribution. Figure 2 presents
experimental results of density-correlation measurements in disordered
lattices. Here, for each realization of disorder, the density auto-correlation
is measured, and then averaged over many realizations (again by shifting
the input site). The result is then normalized, so that $\Gamma_{r,q}=<I_{r}I_{q}>/<I_{r}><I_{q}>$.
As can be seen in Fig 2, both types of disorder show a distinct diagonal
feature in the correlation matrix. The length scale of this feature
along the main diagonal ($q=r$) of the matrix is the localization
length. The width of the diagonal feature is given by the correlation
length, which is not represented in the ensemble-averaged density
distributions. In each single realization, the density distribution
is not a smooth exponentially decaying distribution; it is speckled
\cite{segev,Lahini}. The width of the diagonal feature reflects the
average speckle size. Since in each realization the speckles patter
varies, their featured are smeared out in the averaged distribution,
and the information about their width is lost. However, the fact that
these speckles have a characteristic length scale is recorded in the
averaged correlation function. 

We find that the correlations function carried additional information
on the type of disorder in the lattice, information that is also lost
when one considers the average density distributions. A closer look
at the correlation matrix reveals that for lattices with off-diagonal
disorder (Fig 2c) the correlations tend to form a checkered pattern.
This can be better seen when looking at the correlation function $g(\Delta r)$,
extracted from the correlation matrix $\Gamma_{r,q}$ by summing over
the diagonals, $g(\Delta r)=\sum_{r}\Gamma_{r,r+\Delta r}.$ The density
correlation decays smoothly for lattices with diagonal disorder, yet
it exhibits decaying oscillations for lattices with off-diagonal disorder.
These results were corroborated in numerical simulations (not shown). 

\begin{figure}
\includegraphics[clip,scale=0.99]{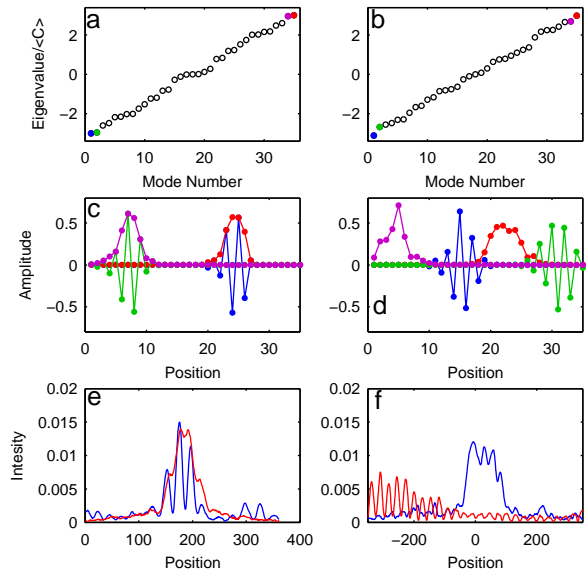}

\caption{\label{fig: 3}(color online). Spectral symmetry in lattices with
diagonal and off-diagonal disorder. (a) The spectrum (band) of eigenvalues
for lattices with off-diagonal disorder. Note the perfect symmetry
with respect to the band center. (b) The spectrum for lattices with
diagonal disorder, showing no such symmetry. (c) Pairs of eigenmodes
taken from symmetric eigenvalues occupy the same region of the lattice,
have the same spatial distribution, but vary in phase structure. (d)
In lattices with diagonal disorder no such symmetry exists. (e) Experimental
measurement of two spectral-symmetric eigenmodes in a lattice with
off-diagonal disorder. (f) A similar measurement in a lattice with
diagonal disorder cannot excite two modes at the same position (see
text). }

\end{figure}

To explain the dependence of the density correlations on the type
of disorder, we start by pointing out a symmetry that exists in periodic
lattices that is sustained also in lattices with pure off-diagonal
disorder, but not in lattices with diagonal disorder\cite{Wegner,Komiyama,Inui,Mudry}.
Without loss of generality, we can set the diagonal terms of the Hamiltonian
to zero, and we denote the randomized tunneling terms as $C_{n}$.
Let $(a_{1},...,a_{N})$ be an eigenvector with an eigenvalue $\lambda$.
Then $C_{1}*a_{2}=\lambda*a_{1}$ ; $C_{1}*a_{1}+C_{2}*a_{3}=\lambda*a_{2}$
; $C_{2}*a_{2}+C_{3}*a_{4}=\lambda*a_{3}$, etc. Now it is obvious
that the vector $b_{n}=(-1)^{n}*a_{n}$, is also an eigenvector with
an eigenvalue $-\lambda$. That is, the eigenvectors are paired around
the center of the band, where twin eigenmodes share the same density
distribution in absolute value, but an opposite (staggered) phase
structure. It is also easy to see that this property will not be exhibited
by lattices with diagonal disorder. For a more detailed proof, see
\cite{Sup}.

To visualize this symmetry, we show in figure 3 the eigenvalues and
eigenmodes for a single realization of a disordered lattice with diagonal
vs. a lattice with off-diagonal disorder. In fig. 3a and 3b, we compare
the spectrum ({}``band'') of eigenvalues for lattices with $N=50$
sites. As was shown for lattices with diagonal disorder \cite{Lahini},
disordered lattices support two types of tightly localized eigenmodes
with eigenvalues at the edges of the spectral band: at one edge the
eigenmodes are tightly localized in space, each mode occupies a different
location, and they generally have a flat phase profile (zero phase
difference between adjacent sites). At the other edge of the spectrum
the modes are also tightly localized, only that they are staggered
- there is a $\pi$ phase difference between adjacent sites. The eigenmodes
at the center of the band have a more complicated phase structure,
and they are typically wider. As we concluded above, in lattices with
off-diagonal disorder we find that each eigenmode at one edge of the
spectrum had a twin-eigenmode at the other edge. These twin eigenmodes
occupy the same region of the lattice, have the same distribution
of the absolute amplitude, but they differ in phase structure as shown
in Fig. 1c. This property does not exist in lattices with diagonal
disorder (see Fig. 1d). 

This special symmetry of the can be observed experimentally. Briefly,
it is possible to excite pure localized eigenmodes by using either
a flat-phase or a staggered beam, with the correct width and initial
position at the lattice. Flat-phased and staggered localized eigenmodes
of a lattice with diagonal disorder were measured (See figure 2 in
\cite{Lahini}), showing indeed that in diagonally disordered lattices
the positions of the lowest-eigenvalues flat-phased eigenmodes never
coincided with the position of the highest-eigenvalues, staggered
localized modes. Fig. 3f shows an excitation of a flat-phased eigenmode
in a lattice with diagonal disorder. Indeed, when the same input beam
was tilted to excite neighboring sites with a $\pi$ phase difference,
the output density showed considerable expansion, suggesting that
no staggered localized eigenmode resides in the same location. In
contrast, Fig 3e, shows the same procedure in a lattice with off-diagonal
disorder. Here, the flat phased beam excited a flat-phased, localized
eigenmode, and a beam with a $\pi$ phase difference between adjacent
sites excited a staggered localized eigenmodes with the same width,
the same spatial profile, and at the same location.

Now, to explain the different density-density correlations shown in
figure 2, we need to consider the effect of these different spectral
properties on the expanding wavepackets when they are excited at a
single site. The localization effect arises from the fact that all
the eigenmodes that are excited have a finite extent. It is well known
that in infinite one dimensional disordered systems all eigenmodes
of the system are localized. In lattices with off-diagonal disorder,
an initial excitation of a single lattice site necessarily involves
the simultaneous excitations of pairs of {}``twin'' eigenmodes\textcolor{black}{,
as they have identical overlap with the initially excited site. The
sum of two identical amplitude distributions yet with a $\pi$-phase
difference in each second site results in a density comb-like pattern
that nulls at every second site. In the dynamic problem several pairs
could be excited simultaneously by the single site initial condition,
and the two modes of the pair accumulate phase in a different rate
(according to their eigenvalues). Nevertheless, the wavepacket will
contain a component with an oscillating intensity pattern, with a
spatial frequency of two-sites. This effect is washed out in the density
distribution averaged over all realizations of disorder, as in each
realization the oscillations appear in a different location. However,
the fact that such oscillations appear in each realization will be
recorded in the averaged correlation.}

\textcolor{black}{We note a previous publication by our group \cite{Qanderson},
in which oscillating }\textcolor{black}{\emph{quantum}}\textcolor{black}{
correlations were analyzed theoretically for bosonic or fermionic
pairs, predicting checker-like correlations in some cases. Those phenomena
are not unrelated to the results reported here, yet the oscillating
correlations reported there were of quantum origin (i.e. in the case
of light they require the use of non-classical light), while here
the described effects are purely classical, wave effects.}

In conclusion, we have experimentally shown that density correlation
measurements can carry a signature of the type of disorder that exists
in a given sample, and we have traced that signature to the existence
of a unique spectral symmetry that is exhibited by lattices with off-diagonal
disorder. Similar results can be measured in matter waves systems,
using density-correlation measurements \cite{Altman} in disordered
lattices. It might also be interesting to study the effect of nonlinear
interactions on these correlations, either in the optical or matter
wave experiments. The signature of these results might also be observed
in correlations measurements for multiply scattered classical \cite{Akkermans}
and non-classical \cite{lodahl,Ott} light. 

This work was supported by the German - Israel Foundation (GIF), the
Minerva Foundation and the Crown Photonics Center. YL acknowledges
support from the Israeli Academy of Science and Humanities.

\end{document}